\documentclass[12pt]{article}
\usepackage{setspace}
\usepackage[utf8]{inputenc}
\usepackage{amsmath}
\usepackage{lineno}
\usepackage{esint}
\usepackage[hidelinks]{hyperref}
\usepackage{graphicx}
\usepackage[english]{babel}
\usepackage[nottoc]{tocbibind}
\usepackage{chngcntr}
\usepackage{array}
\newcolumntype{P}[1]{>{\centering\arraybackslash}p{#1}}
 \usepackage{booktabs}
 \usepackage{tabularx}
 \usepackage{ragged2e} 
\newcolumntype{L}[1]{>{\RaggedRight\arraybackslash}p{#1}} 

\usepackage{float}
\usepackage{subcaption}
\usepackage{algorithm}
\usepackage{forloop}
\usepackage{amsfonts}
\usepackage{algpseudocode}
\usepackage{subcaption}
\usepackage{lineno}
\usepackage[
backend=biber,
style=numeric,
sorting=none
]{biblatex}
\graphicspath{ {./images/} }
\usepackage{geometry}
\newcommand*{\QED}[1][$\square$]{%
\leavevmode\unskip\penalty9999 \hbox{}\nobreak\hfill
    \quad\hbox{#1}%
}
\newtheorem{theorem}{Theorem}
\usepackage{authblk}

\title{\textbf{Dynamic Graph-Based Forecasts of Bookmakers' Odds in Professional Tennis}}

\author[1]{Matthew J. Penn}
\author[2]{Jed Michael}
\author[1,3]{Samir Bhatt\thanks{Corresponding author: \texttt{samir.bhatt@sund.ku.dk}}}

\affil[1]{University of Copenhagen, Copenhagen, Denmark}
\affil[2]{Independent researcher}
\affil[3]{MRC Centre for Global Infectious Disease Analysis, Imperial College London, London, United Kingdom}

\begin{document}
\maketitle
\section*{Abstract}
\textit{Bookmakers' odds consistently provide one of the most accurate methods for predicting the results of professional tennis matches. However, these odds usually only become available shortly before a match takes place, limiting their usefulness as an analysis tool. To ameliorate this issue, we introduce a novel dynamic graph-based model which aims to forecast bookmaker odds for any match on any surface, allowing effective and detailed pre-tournament predictions to be made. By leveraging the high-quality information contained in the odds, our model can keep pace with new innovations in tennis modelling. By analysing major tennis championships from 2024 and 2025, we show that our model achieves comparable accuracy both to the bookmakers and other models in the literature, while significantly outperforming rankings-based predictions.}

\section{Introduction}
Bookmakers' odds have repeatedly been shown to provide some of the most accurate estimates of tennis match outcomes \cite{lisi2017tennis, dryja2025data,fayomi2022forecasting}. However, their usefulness as an analysis tool is limited. Before the start of a tournament, odds are usually only available for first-round matches and for a small number of outright markets (such as a player winning the tournament) \cite{nagel2024price}, which is insufficient to answer every question that may be of interest to a range of stakeholders. For example, a sponsor may be interested in the exact performance of a player in a tournament, enabling them to estimate the value of the exposure that a player may provide \cite{limnell2021professional}. Alternatively, a player may be interested in estimating their expected post-tournament ranking - requiring detailed simulations of multiple players - as this will affect their eligibility to enter subsequent tournaments \cite{reichenberger2019world}. These and similar questions may also be pertinent before the draw has been made, where information from bookmakers' odds is even scarcer.
\\
\\
\noindent
It is therefore important to develop a model which can make predictions for any potential match on any surface. A variety of approaches have already been proposed in the literature, many of which are summarised by Kovalchik  \cite{kovalchik2016searching}. Kovalchik found that attempts to model tennis matches fall into three main categories: regression models, points-based models and paired comparison models. 
\\
\\
\noindent
Regression models \cite{friligkos2023framework, choudhary2023statistical}, use predictors such as the difference in current rank, head-to-head wins and losses, handedness, and height to predict results. These simple models are attractive when limited data is available. Conversely, points-based systems \cite{ingram2019point, spanias2013predicting} are more complex, using estimates of the probability that each player wins a point when serving and returning to then simulate the match. Finally, paired comparison models assign a latent ability that is used to estimate the probability of victory, with variants including Glicko models \cite{yue2022study}, ELO models \cite{angelini2022weighted} and Bradley-Terry models \cite{randrianasolo2022comparing}. In addition to the wealth of established models, the growing poower of machine learning methods means that new approaches are rapidly being developed \cite{sipko2015machine,wilkens2021sports, almarashi2024novel, bayram2021predicting}, meaning that maintaining an industry-leading model requires frequent innovation. Despite this, as previously referenced, there is conclusive evidence in the literature that the bookmakers odds have consistently remained the gold-standard. 
\\
\\
\noindent
The critical observation of this paper is that bookmakers' odds form a compact but rich training set, providing indirect access to industry-leading modelling. This has been previously demonstrated by regression modelling \cite{buhamra2024modeling}, but these approaches can only produce predictions when odds information is already available. Here, we introduce a novel model which, rather than aiming to predict results, instead aims to predict these odds for any match on any surface. This model should be auto-improving, as advances in tennis modelling mean that it has access to increasingly accurate information. Indeed, despite the comparatively small single-feature training dataset, we achieve results comparable to both a range of contemporary models and the bookmakers themselves. 
\\
\\
\noindent
This paper is structured as follows. Firstly, we test our model by applying it to major tennis tournaments in 2024 and 2025, comparing it to the official rankings, the bookmakers, and a range of other models in the literature. Secondly, we present the model methodology. Finally, we discuss the usefulness of our model, and highlight potential areas for improvement.
\section{Results}
Our novel model of forecasting bookmaker odds allows us to assign a time-varying rating to each player which is specific to the surface of interest. We begin by comparing it to both the bookmakers' odds and the official player rankings on a dataset of the most recent (at the time of writing) major championships, before comparing it to a range of models in the literature.
\subsection{Wimbledon 2025 Performance}
To provide a detailed evaluation of our model's behaviour, we test it on the 2025 Wimbledon singles competition, taking data from \url{http://www.tennis-data.co.uk/alldata.php} on both historical results to train the model, and (after the tournament), the results of Wimbledon. We also use the player rankings recorded in these data in our analysis.
\\
\\
\noindent
Note that the restricted coverage of this open source dataset means that there were 25 players in the Wimbledon main draw who had no previous matches in the data. As these players are likely to be very weak, we assign missing players to have the lowest ranking among players competing in the tournament.
\subsubsection{Player Rankings}
\begin{figure}[htbp]
    \centering
    \begin{subfigure}[t]{0.48\textwidth}
        \centering
        \includegraphics[width=\textwidth]{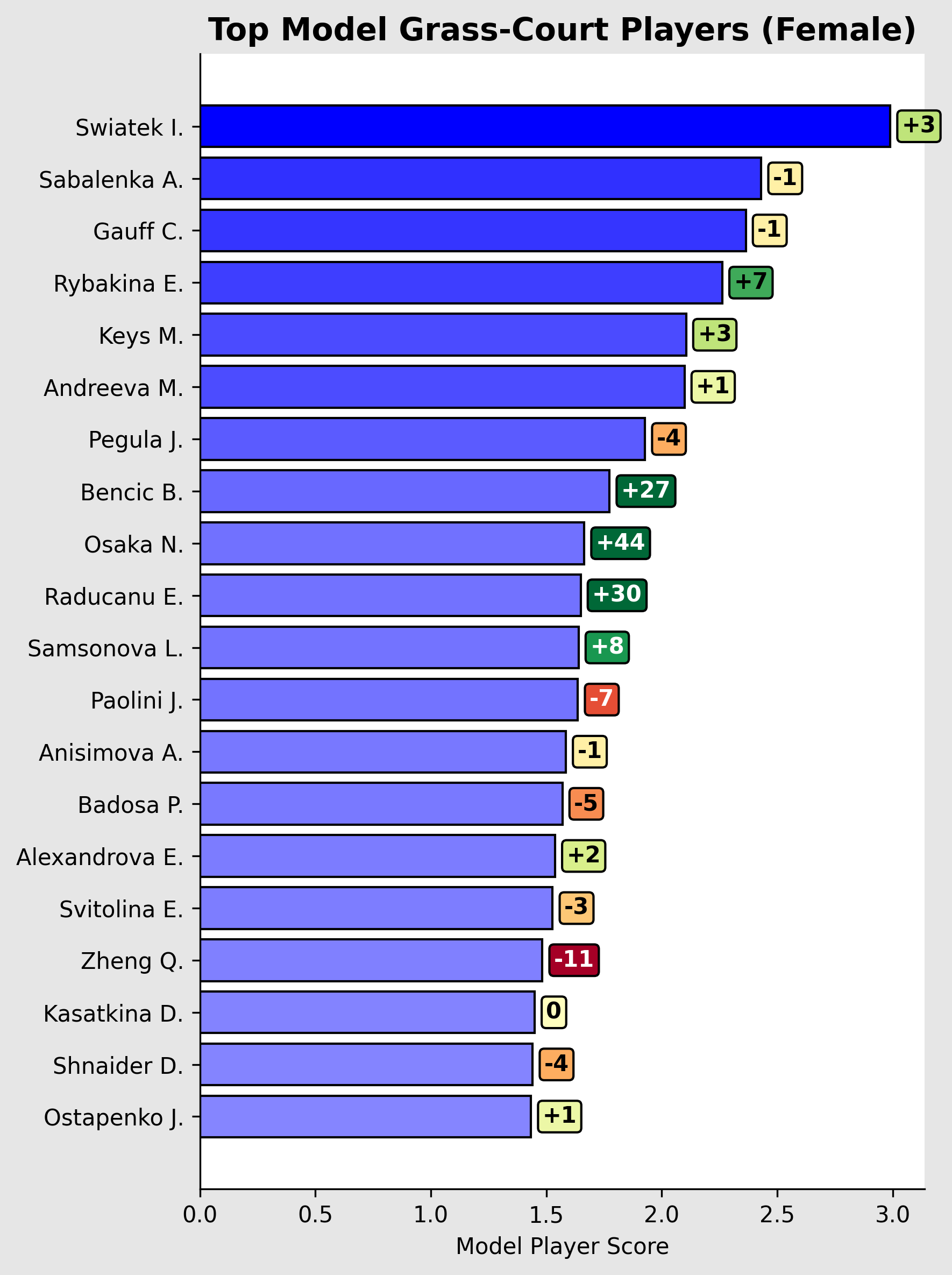}
    \end{subfigure}
    \hfill
    \begin{subfigure}[t]{0.48\textwidth}
        \centering
        \includegraphics[width=\textwidth]{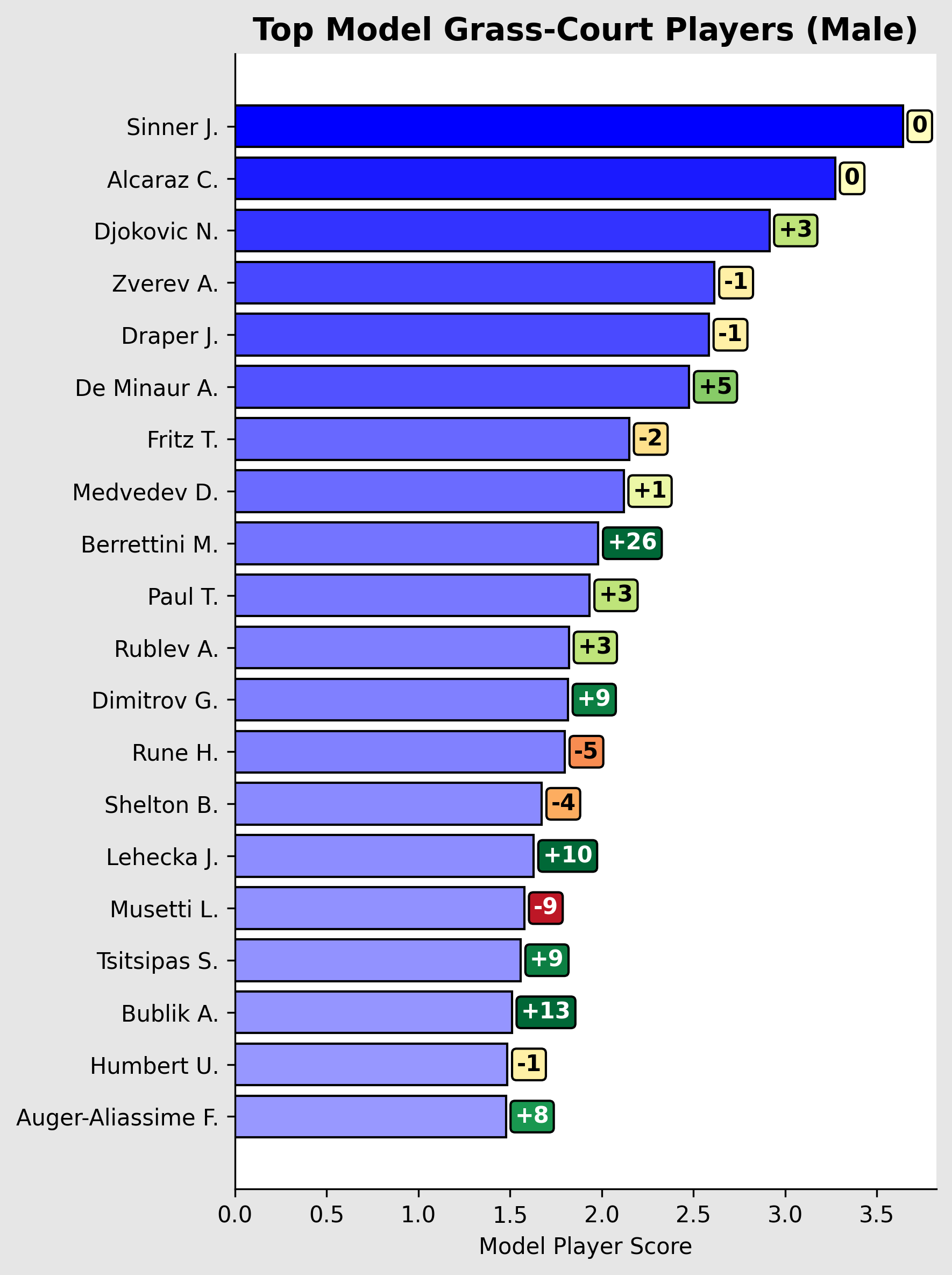}
    \end{subfigure}
    \caption{The top 20 male and female \textbf{grass-court} players at the start of Wimbledon, according to our model. The number at the end of each bar compares the model's ranking to the overall ATP or WTA rankings at that time, with positive numbers indicating that our model ranks that player more highly.}
    \label{fig:overall_ranking}
\end{figure}

 Figure \ref{fig:overall_ranking} shows that our model ranked the eventual winner as its top player in both the men's and women's competition, and ranked all eight semi-finalists within its top 13 players of the relevant gender, notably including Belinda Bencic (officially ranked 35th at the time). In general, its ratings are close to the official rankings (from the ATP and the WTA, for men and women respectively), with 75\% of the players displayed in Figure \ref{fig:overall_ranking} being officially ranked in the top 20).
\\
\\
\noindent
To compare our model to the official rankings, we consider using these rankings to predict results. Discarding the two matches where our model assigned the same ranking to each player (due to the aforementioned data deficiencies), the official rankings predicted 171/252 of results correctly, while our model predicted 182/252. However, this is not a significant result - we will revisit these performance comparisons when we perform a more longitudinal analysis.
\subsubsection{Comparison to Bookmakers' Odds}
\begin{figure}
    \centering
    \includegraphics[width=0.75\linewidth]{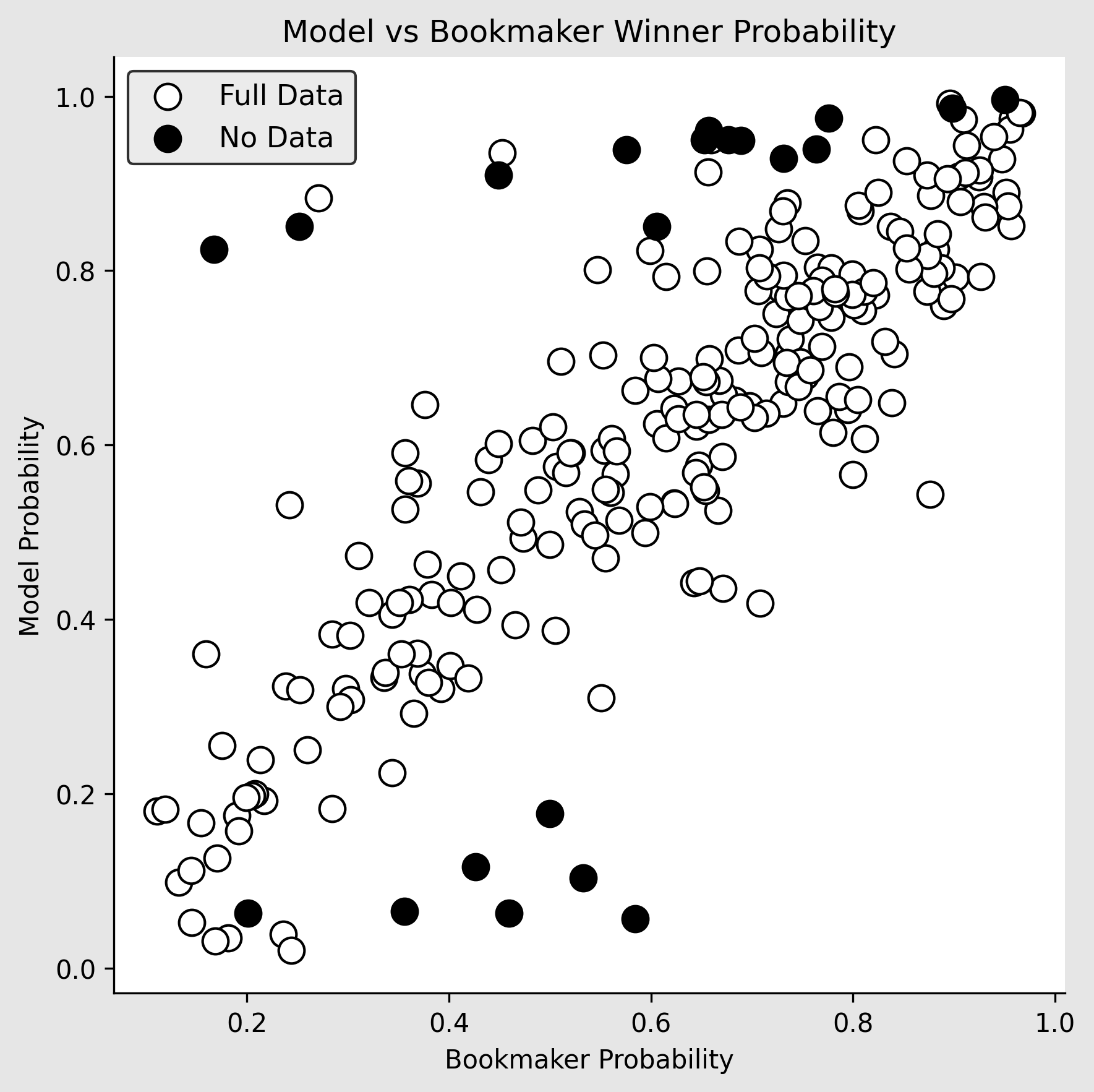}
    \caption{A comparison of the probability assigned to the outcome of each match by the bookmakers and our model. Note that we display the probability assigned to the eventual winner. Black dots show matches where the model had no prior information on at least one of the players. White dots show the remainder of the matches.}
    \label{fig:probability_comparison}
\end{figure}

\begin{figure}
    \centering
    \includegraphics[width=0.75\linewidth]{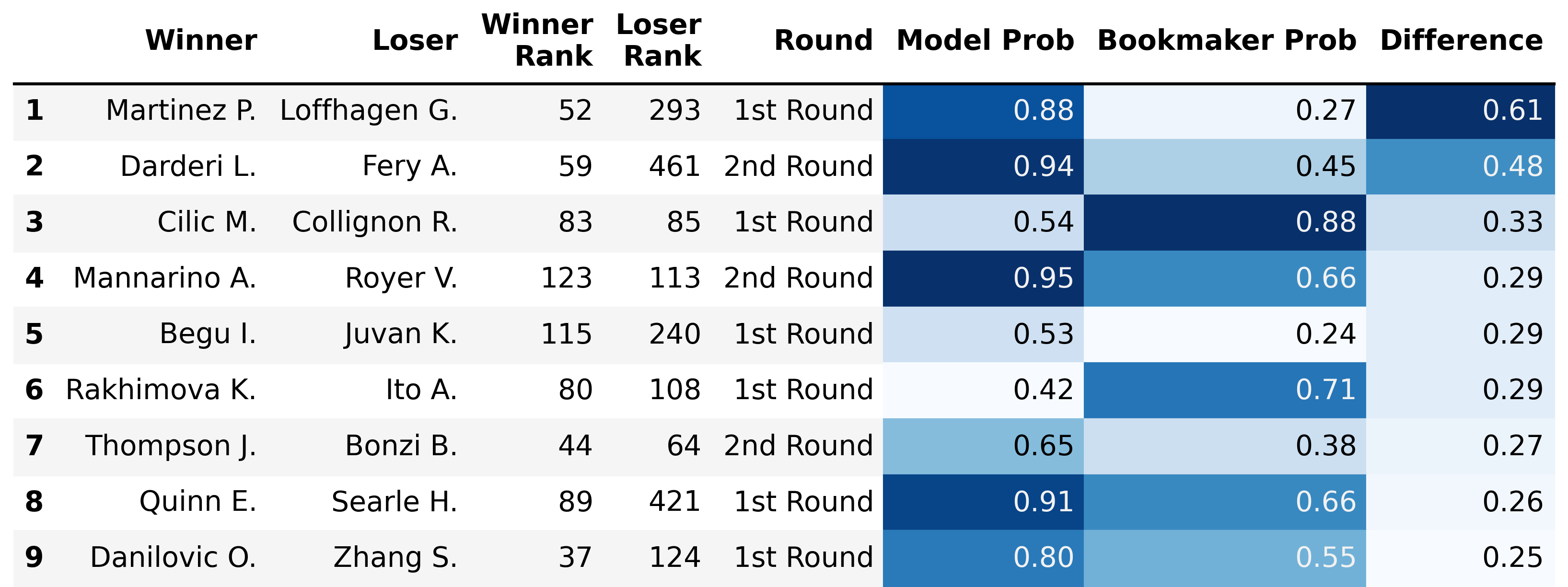}
    \caption{A visualisation of the matches where there was the biggest difference between the model and bookmaker predictions. Note that the `Winner Rank` and `Loser Rank` columns refer to the official ATP/WTA rankings.}
    \label{fig:outliers}
\end{figure}

We compare our model's probabilities to the bookmakers' odds, taken to be the average odds reported in our dataset, which are then normalised through a multiplicative constant to ensure they add to 1. Our model was more successful at forecasting outcomes, though again not significantly so, with the bookmakers predicting only 176/252 results correctly.
\\
\\
\noindent
Figure \ref{fig:probability_comparison} shows a comparison between the forecasts made by the model and the bookmakers. We see immediately that the vast number of the outliers are caused by the model not having data on one of the players. As expected, our model assigns more extreme probabilities to these matches, due to our choice of fixing unknown player abilities to the worst known player in the tournament. This issue would be easily rectified by obtaining a more extensive database of matches. 
\\
\\
\noindent
Restricting attention to matches where data on both players was available, we see that there is a very strong correlation ($\approx 0.88$) between the model and bookmaker forecasts, with a line of best fit close to 1:1 (y = 0.88x + 0.08). This shows that the model is, in general, an accurate tool for predicting the bookmakers' odds ahead of them being released.
\\
\\
\noindent
It is informative to examine the other outliers (where data was available to our model), which are displayed in Figure \ref{fig:outliers}. The largest two (Martinez v Loffhagen and Darderi v Fery), and also Quinn v Searle, are notable for containing British players playing against, at least according to the official rankings, much stronger opponents. This raises two major points for discussion. Firstly, there is no mechanism for home advantage within the model. Certainly, in other sports \cite{gomez2011comparison}, this is a major factor and is something that would be beneficial to include in a future iteration of the model. However, it seems unlikely that home advantage alone can explain the dramatic swings seen here. Instead, it seems likely that there is evidence of either the bookmakers hedging (due to strong support for the British players) or perhaps trying to encourage bets from the British public on these players, with their quoted odds perhaps appearing more palatable to the average punter. Certainly, it seems that these odds are not a true reflection of the match outcome probability, and that our model can, in these cases, provide more robust, and certainly more explainable forecasts.
\\
\\
\noindent
The low odds given by the bookies to Thompson against Bonzi can be explained by Thompson's withdrawal through injury from the HSBC Championships \cite{atp2025stats} less than two weeks before the start of Wimbledon. Such information is not included in our model, and therefore it can act as a detection system for possible player injuries when, as in this case, the bookmakers' odds are lower than expected.
\\
\\
\noindent
The remaining outliers from Figure \ref{fig:outliers} appear to be due to low data volume on the players involved - in particular, since the start of 2025, our dataset included four matches from Zhang and Collignon, three from Ito, and only one from Royer and Juvan (which compares to, for example, 51 matches from Sabalenka). This is clearly a far less interesting or useful property of the model, but could be fixed through the acquisition of more comprehensive data. 
\subsection{Longitudinal Performance}
\begin{figure}
    \centering
    \includegraphics[width=0.9\linewidth]{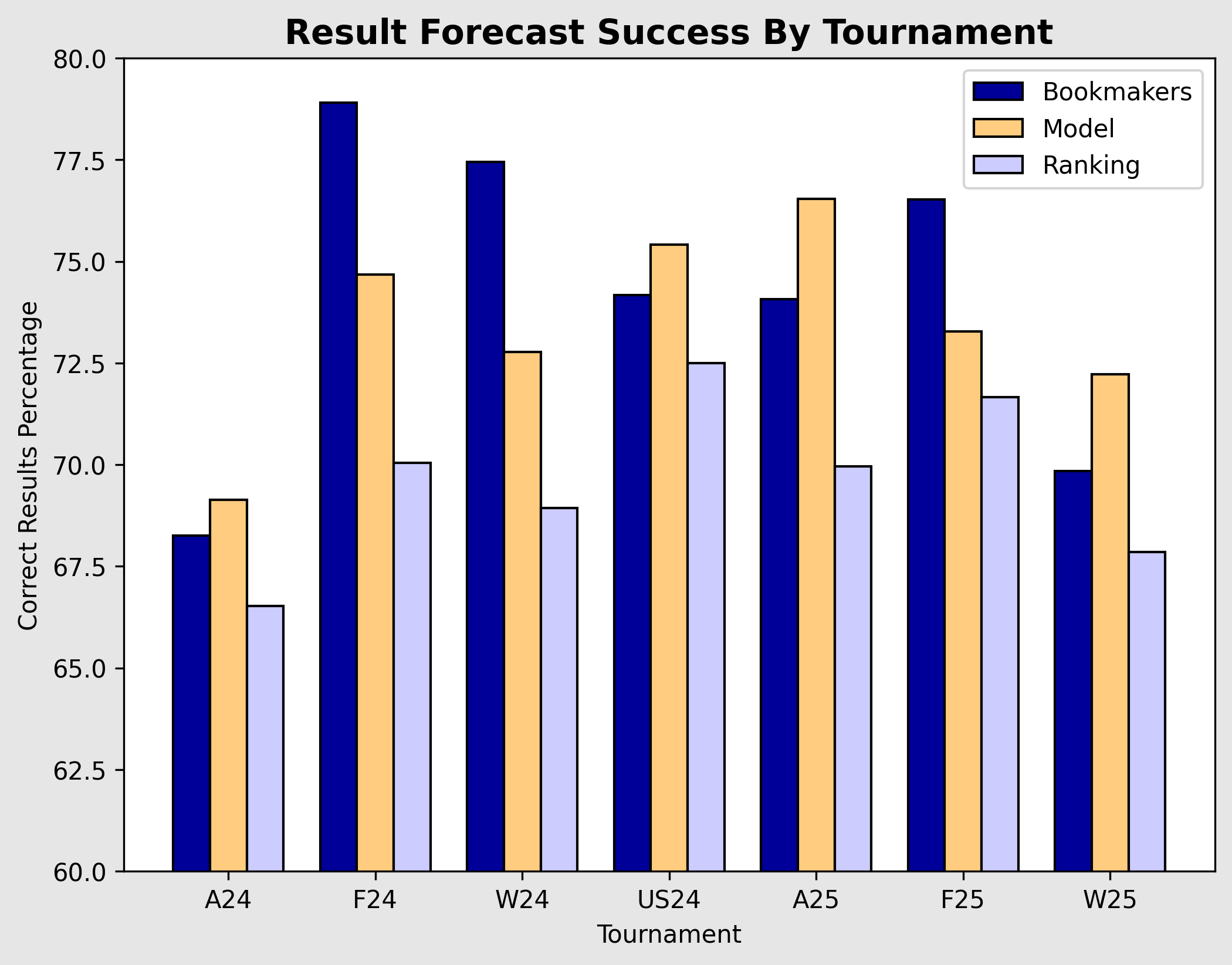}
    \caption{The percentage of results that were predicted correctly by the model, the bookmakers and the ATP/WTA rankings in each of the majors (note that A = Australian Open, F = French Open, W = Wimbledon and US = US Open). Note further that the y-axis begins at 60\% to highlight the differences between the three predictions.}
    \label{fig:results_by_tournament}
\end{figure}

While it is informative to examine a single tournament closely, obtaining statistically significant results is difficult due to the relatively small number of matches, and the general agreement between predictions from our model, the bookmakers, and the rankings.
\\
\\
\noindent
To increase the available data, we therefore use our model to predict the major championships\footnote{That is, the Australian Open, the French Open, Wimbledon, and the US Open} from 2024 and 2025 (note that at the time of writing, the 2025 US Open had not happened, and so this leaves seven major championships to be forecast).
\\
\\
\noindent
Figure \ref{fig:results_by_tournament} shows the performance of the model across these seven tournaments. Our model predicted more results correctly than the official rankings in each of the seven tournaments, while outperforming the bookmakers in four of these. Note that again, matches where the model assigned the same ranking to each player have been discarded (on average, 13 per tournament - with more in historical tournaments due to more limited data coverage).
\\
\\
\noindent
It is also informative to consider the overall number of results that were predicted correctly. Out of the 1684 total matches in our dataset, the bookmakers predicted 1249 (74.1\%) correctly, our model predicted 1237 (73.5\%), and the rankings predicted 1173 (69.7\%). Performing a simple two-sided hypothesis test, this means that there was no significant difference between our model and the bookmakers (p = 0.40), but there was a significant difference between our model and the rankings (p = 0.0006).
\subsection{Comparison to Models in the Literature}
In order to understand how our model's performance compares to its contemporaries, it is important to compare its predictive success with those reported in the literature. However, such a task in non-trivial as (as demonstrated in Figure \ref{fig:results_by_tournament}) the predictability of tournaments can vary substantially. Thus, without training and testing each model on the same datasets, one cannot use the raw outputs, such as predictive accuracy, to fairly compare the ability of different models.
\\
\\
\noindent
Instead, we suppose that the bookmakers offer a gold-standard benchmark, and therefore simply assess the performance of other models in comparison to the bookmakers (which is often reported as a baseline for the specific tournaments that each model was trained on). We define the following two scores:
\begin{align}
    \text{Ratio} &= 100\times\bigg[\frac{\text{Model Accuracy}}{\text{Bookmakers' Accuracy}} - 1\bigg] \\
    \text{Difference} &= 100\times \bigg[(\text{Model Accuracy}) - (\text{Bookmakers' Accuracy})\bigg]
\end{align}
In both cases, a score of 0 indicates equivalent performance to the bookmakers, while positive scores indicate better performance. Note that we give each score as a percentage.

\begin{table}[htbp]
    \centering
    \begin{tabularx}{\textwidth}{L{6cm} c c c c}
        \toprule
        \textbf{Model} & \textbf{Matches} & \textbf{Date Range} & \textbf{Ratio} & \textbf{Difference} \\
        \midrule
        Gradient Boosting (including bookmaker odds) \cite{wilkens2021sports} & 3996 & 2013--2019 & $+0.14$ & $+0.10$ \\
        Random Forest \cite{dryja2025data} & 7620 & 2010--2024 & $-0.11$ & $-0.14$ \\
        Bradley-Terry \cite{fayomi2022forecasting} & 3439 & 2019--2020 & $-0.47$ & $-0.32$ \\
        \textbf{This Paper} & 1684 & 2024--2025 & $-0.96$ & $-0.71$ \\
        Logistic Regression \cite{lisi2017tennis} & 501 & 2013 & $-1.03$ & $-0.80$ \\
        ELO \cite{kovalchik2016searching} & 2395 & 2014 & $-2.78$ & $-2.00$ \\
        Points-Based \cite{kovalchik2016searching} & 2395 &2014 & $-6.94$ & $-5.00$ \\
        \bottomrule
    \end{tabularx}
    \caption{Comparison of model performance against the bookmakers. Note that where multiple similar models were considered in a paper, only the best-performing model has been presented.}
    \label{tab:model_comp}
\end{table}
\noindent
Using these comparative measures, Table \ref{tab:model_comp} shows that our model performs averagely well compared to those found in the literature. It should be noted that a number of the models in Table \ref{tab:model_comp} are evaluated on relatively old data, and therefore, one would postulate that they are competing against worse bookmaker models. It is notable that the best-performing model \cite{wilkens2021sports} surveyed uses the bookmaker odds as part of its training process, and that it only managed a small improvement on the bookmaker benchmark, again highlighting the power of this data. Certainly, our model is competitive with its contemporaries, and can be a useful tool for match prediction.
\section{Methods}
We suppose that we have access to bookmakers' odds $\boldsymbol{d}$ on historical tennis matches. From these odds, we can estimate the bookmakers' true probabilities  $\boldsymbol{p}$ for each outcome, which we do by setting
\begin{equation}
    \boldsymbol{p} = \frac{\boldsymbol{d}}{||\boldsymbol{d}||_1}
\end{equation}
\subsection*{Setup}
Our model starts with the assumption that the bookmakers' odds are based an approximate, ELO-style model, with player ratings that may depend on the surface that the match is played on. Given ratings $r_a$ and $r_b$ for players $a$ and $b$, this therefore means that 
\begin{equation}
\label{eq:elo_eq}
    \mathbb{P}(\text{$a$ beats $b$}) = \frac{1}{1 + 10^{(r_b - r_a)}} := p_{ab}
\end{equation}
Importantly, for the rest of our model definition, we observe that if we define the quantities
\begin{equation}
\label{eq:elo_add}
    x_{ab} := \log\bigg(\frac{1-p_{ab}}{p_{ab}}\bigg) = r_a - r_b
\end{equation}
then we have, for any $c$
\begin{equation}
    x_{ab} = r_a - r_b =(r_a - r_c) + (r_c - r_b) = x_{ac} + x_{cb}
\end{equation}
For the remainder of this paper, we will call $x_{ab}$ the \textbf{log-transformed odds}.
\\
\\
\noindent
Note that we are not assuming that the bookmakers use a true ELO model (as it appears almost certain that their modelling process is far more complex). We assume only that they somehow assign players a dynamic rating, and that the match results probabilities can be estimated using (\ref{eq:elo_eq}).
\subsection{Three- and Five-Set Matches}
While the majority of professional tennis matches are best-of-three sets, the four grand slams on the men's tour are instead best-of-five. This increased number of sets reduces the variance of the outcome, and must be accounted for in our model.
\\
\\
\noindent
To do this, we use the simplifying assumption that sets are independent, and can therefore fit a set win probability $\xi$ by solving the equation
\begin{equation}
    \mathbb{P}\left(\text{Bin}(N_s, \xi) > \frac{N_s}{2}\right) = p_{ab}
\end{equation}
This allows us to impute three-set win probabilities from the bookmakers' odds for five-set matches where appropriate.
\subsection{A Graphical Structure for Log-Transformed Odds}
Given the above observation, we now introduce a directional, weighted, complete graph $(\mathcal{G},E)$ where each node represents a player and $E_{ab}$ is the distance of the edge between nodes $a$ and $b$. Our aim is that this graph represents the log-transformed odds, such that $E_{ab}$ is the log-transformed odds of player $a$ beating player $b$.
\\
\\
\noindent
We use historical match odds to choose the weights on this graph. Given a single match between players $a$ and $b$ and the bookmakers' log-transformed odds $x_{ab}$, the natural approach would be to set
\begin{equation}
    E_{ab} = x_{ab}
\end{equation}
However, this neglects the importance of both the time in the past, and the surface on which the match was played. For a match $M$, played $t_M$ days in the past on surface $s$, we define the weight $w_{ab}(M)$ to be
\begin{equation}
    w_{ab}(M) := \rho^{t_M} \tau_{s}
\end{equation}
 where $\rho$ controls the strength of our temporal decay and $\tau_s$ is the surface-specific weight. 
\\
\\
\noindent
Defining $W_{ab}$ to be the total weight assigned to an edge $(a,b)$, we then have
\begin{align}
    W_{ab}  &= \sum_M w_{ab}(M) \\
    E_{ab} &= \frac{\sum_{M}w_{ab}(M)x_{ab}(M)}{W_{ab}}
\end{align}
\\
\\
\noindent
 The choice of geometric temporal decay may appear arbitrary, but it is essentially a Markovian assumption on our player ability - namely, the assumption that our estimation of player ability tomorrow requires only our estimation today (and any new matches that occur tomorrow!). A proof that this assumption results in the above geometric decay is given in the appendix.
\\
\\
\noindent
Computational efficiency arises from the fact that, if the last match between players $a$ and $b$ was at time $t$ in the past and we process a new match $M$ today, we can simply set
\begin{align}
     W'_{ab}  &= \rho^{t}W_{ab} + w_{ab}(M) \\
    E'_{ab} &= \frac{W_{ab}\rho^{t}E_{ab} + w_{ab}(M)x_{ab}(M)}{W'_{ab}}
\end{align}
where $'$ denotes our old quantities\footnote{Note that in practice, it is simpler to keep track of $W_{ab}E_{ab}$ instead of $E_{ab}$, as we do in our code, and only normalise when necessary}. In particular, as suggested by our Markovian principle, we do not need to maintain a history of all previous matches and their weights, allowing our method to scale to large datasets.
\subsection{Fitting Player Ranks}
We aim to find the player ranks $\boldsymbol{r}$ which ``best approximate'' the graph defined by $E$. We define our objective function to be the weighted sum of squares
\begin{equation}
    f(\boldsymbol{r}) = \sum_{a,b}W_{ab}\bigg((r_a - r_b) - E_{ab}\bigg)^2
\end{equation}
Note that for $a\neq b$
\begin{equation}
    \frac{\partial f}{\partial r_a \partial r_b}= -2(W_{ab} + W_{ba})
\end{equation}
while
\begin{equation}
    \frac{\partial^2 f}{\partial r_a^2} = 2\sum_{b: a \neq b}(W_{ab} + W_{ba})
\end{equation}
In particular, this means that the Hessian $H$ satisfies
\begin{equation}
    \sum_{b:a\neq b} |H_{ab}| = 2\sum_{b: a \neq b}(W_{ab} + W_{ba}) = H_{ab}
\end{equation}
and hence, by Gershgorin's Theorem \cite{horn2012matrix}, $H$ is positive semidefinite. This means that $f$ is convex, and therefore, an optimal player ranking can be easily defined.
\\
\\
\noindent
Note that there is clearly a degree of freedom in the optimal solution, as adding a constant to each player ranking leaves the $f$ unchanged. Moreover, one must take care in the case where there are multiple disjoint groups of players (that is, if one removes any edges from $\mathcal{G}$ with weight zero, then there are multiple disconnected subgraphs of $\mathcal{G}$). This is not practically a problem for predicting results in major tennis tournaments, but could be important for forecasting odds in competitions involving players with few past results in the dataset. In this case, adding a prior on player ability could help to align the disjoint clusters.
\subsection{Implementation Notes}
To fit the weights $\tau_s$ for each surface, we performed a grid-search, evaluating on historical tournaments within the dataset to find the optimal prediction accuracy. We noted relatively insensitive dependence on these parameters, and therefore did not fit these parameters exactly - in part due to the computational requirements of performing each evaluation (approximately 5 minutes per evaluation).
\\
\\
\noindent
The methods presented here are implemented in a Python 3.11 repository available here \url{https://github.com/mpenn114/rss-wimbledon-2025}. Data was taken from \url{http://www.tennis-data.co.uk/alldata.php}, as explained in the repository. Optimisation was performed using the package \textit{scipy} \cite{virtanen2020scipy} using the \textit{L-BFGS-B} solver for efficiency. The model was run on a Lenovo Thinkpad with quad-core 2.7GHz Intel i7 processors and 32GB of RAM. Training player strengths for a single tournament took approximately 5 minutes, though this can be reduced by decreasing the maximum number of iterations in the strength-fitting algorithm.
\section{Discussion}
The novel forecasting model introduced in this paper is an effective match prediction tool with a similar success rate to bookmakers. Across 1684 Grand Slam matches, our model predicted 1237 (73.5\%) correctly, with the gap between it and the bookmakers being comparable to other models in the literature. This success comes despite its relatively small training dataset, which makes it possible to realistically apply it on open-source data. 
\\
\\
\noindent
The complete reliance of our training dataset on the bookmakers' odds means that (with the exception of matches where the bookmakers have a commercial incentive to list odds which are not equal to their best prediction of match outcome) our model cannot be expected to outperform the bookmakers. However, as has been repeatedly shown in the literature, the bookmakers' models have consistently been the gold standard. Indeed, this is almost necessarily true - were there a model in the literature which significantly outperformed the bookmakers, it would be in the bookmakers' interest to develop a similar approach, as otherwise they would be vulnerable to making substantial losses. Thus, a model that has comparable accuracy to the bookmakers is likely to provide predictions that are competitive in accuracy with any other publicly-available method.
\\
\\
\noindent
As well as its current strong performance, our model has the potential to report improved results - in absolute terms - as the bookmakers' odds themselves improve, though it may decline in relative performance as the more complex underlying bookmakers' models become less well-approximated by our ELO assumption. This is difficult to test empirically as the ``difficulty'' of tournament prediction varies over time due to the differing distribution of player abilities in the draw. However, our model certainly appears to have the most potential to be future-proof in comparison to those found in the literature which do not use odds-related information. 
\\
\\
\noindent
Alongside its obvious application as a prediction tool, our model also has the ability to identify anomalies in the bookmakers' odds. As shown during Wimbledon 2025, there were major outliers which could be explained through either commercial reasons, or external factors such as player injuries. This information can be useful both for those seeking to place bets on matches, or other stakeholders - such as media outlets - who could then investigate the reasons for these anomalies further.
\\
\\
\noindent
Moreover, it should be possible to apply this model to a range of other sports. Certainly, the method would need very little adjustment to be used in other individual sports with binary outcomes and frequent matches, such as, for example, darts or snooker. In addition, provided one could develop similar rating equations to (\ref{eq:elo_add}), it should be possible to apply this framework more generally - for example, by using an assumed underlying Double Poisson model \cite{penn2022analysis} in football.
\\
\\
\noindent
As well as widening its scope, there are, of course, areas in which this model could improve. For example, as shown in \cite{wilkens2021sports}, combining the bookmakers' odds with other features, such as information on actual match outcomes, can provide improved results. This could help to mitigate the impact of anomalous odds, and could provide more explainability to the outputs of the models. Such anomalous predictions, where either a player is injured or the profit-maximising odds for the bookmaker differ from their best estimate of the probabilities, can be problematic in our training dataset, causing our own player ratings to deviate from the bookmakers'. Having other features could allow us to, for example, have confidence in discarding data that we think is not representative of overall player ability.
\\
\\
\noindent
Moreover, our model would benefit from a more sensible method of assigning player abilities when no previous odds information was information. Figure \ref{fig:outliers} showed that our predictions diverged substantially from the bookmakers' odds in this case, and we saw further that this could occur when only a small number of past matches were available. Even a simple method based on current official ranking could help to reduce this discrepancy and enable our model to provide more robust predictions in these cases.  
\section{Conclusion}
\begin{itemize}
    \item We have developed a model based on historical bookmakers' odds that aims to forecast these odds for future matches.
    \item This model dynamically adjusts its player ratings over time as new odds information becomes available, and can provide surface-specific predictions.
    \item Our model performed strongly in major tennis championships from 2024 and 2025, achieving a level of accuracy that was only slightly lower than the bookmaker model.
    \item The gap between our model and the bookmakers' is comparable to other models found in the literature.
\end{itemize}
\section*{Data Availability}
The methods presented here are implemented in a Python 3.11 repository available here \url{https://github.com/mpenn114/rss-wimbledon-2025}. Data was taken from \url{http://www.tennis-data.co.uk/alldata.php}, as explained in the repository. 
\printbibliography

@book{horn2012matrix,
  title={Matrix analysis},
  author={Horn, Roger A and Johnson, Charles R},
  year={2012},
  publisher={Cambridge university press}
}

@misc{atp2025stats,
  author       = {{ATP Tour}},
  title        = {Match Stats Archive – Miami 2025},
  year         = {2025},
  url          = {https://www.atptour.com/en/scores/stats-centre/archive/2025/311/ms017},
  note         = {Accessed: 2025-07-29},
  howpublished = {\url{https://www.atptour.com/en/scores/stats-centre/archive/2025/311/ms017}}
}

@article{virtanen2020scipy,
  title={SciPy 1.0: fundamental algorithms for scientific computing in Python},
  author={Virtanen, Pauli and Gommers, Ralf and Oliphant, Travis E and Haberland, Matt and Reddy, Tyler and Cournapeau, David and Burovski, Evgeni and Peterson, Pearu and Weckesser, Warren and Bright, Jonathan and others},
  journal={Nature methods},
  volume={17},
  number={3},
  pages={261--272},
  year={2020},
  publisher={Nature Publishing Group US New York}
}

@article{gomez2011comparison,
  title={Comparison of the home advantage in nine different professional team sports in Spain},
  author={G{\'o}mez, Miguel A and Pollard, Richard and Luis-Pascual, Juan-Carlos},
  journal={Perceptual and motor skills},
  volume={113},
  number={1},
  pages={150--156},
  year={2011},
  publisher={SAGE Publications Sage CA: Los Angeles, CA}
}

@article{lisi2017tennis,
  title={Tennis betting: can statistics beat bookmakers?},
  author={Lisi, Francesco and Zanella, Germano},
  journal={Electronic Journal of Applied Statistical Analysis},
  volume={10},
  number={3},
  year={2017}
}

@phdthesis{dryja2025data,
  title={Data-Driven Prediction of ATP Tennis Match Outcomes Using Machine Learning Techniques},
  author={Dryja, Jakub Olaf},
  year={2025},
  school={Vrije Universiteit Amsterdam}
}

@article{fayomi2022forecasting,
  title={Forecasting Tennis Match Results Using the Bradley-Terry Model},
  author={Fayomi, Aisha and Majeed, Rizwana and Algarni, Ali and Akhtar, Sohail and Jamal, Farrukh and Nasir, Jamal Abdul},
  journal={International Journal of Photoenergy},
  volume={2022},
  number={1},
  pages={1898132},
  year={2022},
  publisher={Wiley Online Library}
}

@article{nagel2024price,
  title={Price Formation Dynamics and Learning in the Tennis Sports Betting Market},
  author={Nagel, Michael},
  journal={Available at SSRN 5043714},
  year={2024}
}

@article{limnell2021professional,
  title={Professional players, their brands, and the competitive strategies of tennis clothing companies},
  author={Limnell, Ville},
  year={2021}
}

@phdthesis{reichenberger2019world,
  title={World ranking points as incentive-Does the awarding of ATP points affect tennis players' willingness to participate in the Davis Cup?},
  author={Reichenberger, Clemens},
  year={2019},
  school={University of Salzburg}
}

@article{kovalchik2016searching,
  title={Searching for the GOAT of tennis win prediction},
  author={Kovalchik, Stephanie Ann},
  journal={Journal of Quantitative Analysis in Sports},
  volume={12},
  number={3},
  pages={127--138},
  year={2016},
  publisher={De Gruyter}
}

@article{friligkos2023framework,
  title={A framework for applying the Logistic Regression model to obtain predictive analytics for tennis matches.},
  author={Friligkos, Georgios and Papaioannou, Evi and Kaklamanis, Christos},
  journal={Technium},
  volume={15},
  year={2023}
}

@article{choudhary2023statistical,
  title={A statistical model to predict the results of Novak Djokovic’s matches in the Australian open tennis event using the binary logistic regression},
  author={Choudhary, Prashant Kumar and Dubey, Suchishrava and Brijwal, Dinesh and Paswan, Rajan},
  journal={International Journal of Statistics and Applied Mathematics},
  volume={8},
  number={1},
  pages={17--21},
  year={2023}
}

@article{ingram2019point,
  title={A point-based Bayesian hierarchical model to predict the outcome of tennis matches},
  author={Ingram, Martin},
  journal={Journal of Quantitative Analysis in Sports},
  volume={15},
  number={4},
  pages={313--325},
  year={2019},
  publisher={De Gruyter}
}

@article{spanias2013predicting,
  title={Predicting the outcomes of tennis matches using a low-level point model},
  author={Spanias, Demetris and Knottenbelt, William J},
  journal={IMA Journal of Management Mathematics},
  volume={24},
  number={3},
  pages={311--320},
  year={2013},
  publisher={OUP}
}

@article{yue2022study,
  title={A study of forecasting tennis matches via the Glicko model},
  author={Yue, Jack C and Chou, Elizabeth P and Hsieh, Ming-Hui and Hsiao, Li-Chen},
  journal={Plos one},
  volume={17},
  number={4},
  pages={e0266838},
  year={2022},
  publisher={Public Library of Science San Francisco, CA USA}
}

@article{angelini2022weighted,
  title={Weighted Elo rating for tennis match predictions},
  author={Angelini, Giovanni and Candila, Vincenzo and De Angelis, Luca},
  journal={European Journal of Operational Research},
  volume={297},
  number={1},
  pages={120--132},
  year={2022},
  publisher={Elsevier}
}

@inproceedings{randrianasolo2022comparing,
  title={Comparing Different Data Representations and Machine Learning Models to Predict Tennis},
  author={Randrianasolo, Arisoa S and Pyeatt, Larry D},
  booktitle={Future of Information and Communication Conference},
  pages={488--500},
  year={2022},
  organization={Springer}
}

@article{wilkens2021sports,
  title={Sports prediction and betting models in the machine learning age: The case of tennis},
  author={Wilkens, Sascha},
  journal={Journal of Sports Analytics},
  volume={7},
  number={2},
  pages={99--117},
  year={2021},
  publisher={SAGE Publications Sage UK: London, England}
}

@article{almarashi2024novel,
  title={A novel comparative study of NNAR approach with linear stochastic time series models in predicting tennis player's performance},
  author={Almarashi, Abdullah M and Daniyal, Muhammad and Jamal, Farrukh},
  journal={BMC Sports Science, Medicine and Rehabilitation},
  volume={16},
  number={1},
  pages={28},
  year={2024},
  publisher={Springer}
}

@inproceedings{bayram2021predicting,
  title={Predicting tennis match outcomes with network analysis and machine learning},
  author={Bayram, Firas and Garbarino, Davide and Barla, Annalisa},
  booktitle={International Conference on Current Trends in Theory and Practice of Informatics},
  pages={505--518},
  year={2021},
  organization={Springer}
}

@article{sipko2015machine,
  title={Machine learning for the prediction of professional tennis matches},
  author={Sipko, Michal and Knottenbelt, William},
  journal={MEng computing-final year project, Imperial College London},
  volume={2},
  year={2015}
}

@article{buhamra2024modeling,
  title={Modeling and prediction of tennis matches at Grand Slam tournaments},
  author={Buhamra, N and Groll, A and Brunner, S},
  journal={Journal of Sports Analytics},
  volume={10},
  number={1},
  pages={17--33},
  year={2024},
  publisher={SAGE Publications Sage UK: London, England}
}

@article{penn2022analysis,
  title={Analysis of a double Poisson model for predicting football results in Euro 2020},
  author={Penn, Matthew J and Donnelly, Christl A},
  journal={Plos one},
  volume={17},
  number={5},
  pages={e0268511},
  year={2022},
  publisher={Public Library of Science San Francisco, CA USA}
}
\section*{Appendix}
In this appendix, we prove the claim made in the main text about our choice of weights to ensure that we require only our current estimation for player ability. Note that we assume in this theorem that $w$ is differentiable to reduce its complexity, though believe it should hold for less stringent conditions. We also assume that there is a single weight function dependent only on time (excluding, for example, surface-dependent weights). Again, this proof could be extended to include these considerations.
\begin{theorem}
    Suppose that an edge weight $E(0)$ at time $0$ is given by
    \begin{equation}
        E(0) = \frac{\sum_M w(t_M)x(M)}{\sum_Mw(t_M)}
    \end{equation}
    for some differentiable weight function $w(t_M) > 0$, which is the weight given to match $M$ which occurred at $t_M$ time in the past.
    \\
    \\
    \noindent
    We suppose that if we observe any new match at time $t^* > 0$ in the future with log-transformed odds $x^*$, then, for any initial set of matches and values of $x^*$ and $t^*$, the new edge weight, $E(t^*)$, is given by some function $F$ that depends only on $E(0)$ and the new match. That is
    \begin{equation}
        E(t^*) = F(E(0), t^*, x^*)
    \end{equation}
    Then, it is necessary that
    \begin{equation}
        w(t_M) = A\rho^{t_M}
    \end{equation}
    for some constants $A\geq0$ and $\rho\geq0$.
\end{theorem}
\textbf{Proof:} Note that we can express $E(t^*)$ as
\begin{equation}
        E(t^*) = \frac{\sum_M w(t_M + t^*)x(M) + x^*w(0)}{\sum_Mw(t_M + t^*) + w(0)}
    \end{equation}
From the statement of the theorem, we need only show that the stated form of $w$ is necessary in one specific setup. To this end, to simplify the problem, we suppose that we have two matches in the past, at times $0$ and $-t$, with the new match being at time $t$. We represent these matches as $M_1$, $M_2$, and $M_3$ respectively. Furthermore, noting that multiplying all the weights by a constant leaves our expressions for $E$ unchanged, we set $w(t) + w(0) = 1$ This then simplifies our equation to
\begin{equation}
    E(0) = w(t)x(M_1) + w(0)x(M_2)
\end{equation}
Now, having only observed $E(0)$, it makes sense to transform our variable system $(x(M_1), x(M_2))$ to $(E(0), k)$ where we have
\begin{equation}
    x(M_2) = k \quad \text{and} \quad x(M_1) = \frac{1}{w(t)}\bigg[E(0) - w(0)k\bigg]
\end{equation}
In this system, we can freely change $k$ to be any value, and therefore require that $E(1)$ is independent of $k$. However,
\begin{align}
    E(1) &= \frac{ w(2t)x(M_1) + w(t)x(M_2) + w(0)x(M_3)}{1 + w(2t)} \\
    &=\frac{ \frac{w(2t)}{w(t)}\bigg[E(0) -w(0)k\bigg]  + w(t)k + w(0)x(M_3)}{1 + w(2t)} \\
    &= \frac{w(2t)E(0)}{w(t)(1+w(2t))} + \frac{w(0)x(M_3)}{1+w(2t)} + \frac{\bigg[w(t) - \frac{w(2t)w(0)}{w(t)}\bigg]k}{1+w(2t)} \\
\end{align}
This expresses the equation for $E(1)$ as the sum of three terms, the first of which depends only on $E(0)$ and $w$, the second of which depends only on the new match $x(M_3)$ and $w$, and the third of which depends on our free parameter $k$. For $E(1)$ to be independent of $k$, it is therefore necessary that
\begin{equation}
    w(t) - \frac{w(2t)w(0)}{w(t)} = 0 \Rightarrow w(t)^2 = w(0)w(2t)
\end{equation}
Now, define the function $v(t) = \log(w(t))$ (which is valid as $w(t) > 0$). This gives
\begin{equation}
    2v(t) = v(2t) + v(0)
\end{equation}
which implies
\begin{equation}
    2(v(t) - v(0)) = v(2t) - v(0)
\end{equation}
Thus, we further simplify by defining $u(t) = v(t) - v(0)$ so that we have
\begin{equation}
    u(2t) = 2u(t) 
\end{equation}
and therefore, inductively, for any $n$,
\begin{equation}
    u(2t) = 2^nu\bigg(\frac{t}{2^{n-1}}\bigg)
\end{equation}
Now, we know that $u$ is differentiable as it is a continuous function of the differentiable $w$, and therefore, noting further that $u(0) = 0$, we see that
\begin{equation}
   \lim_{n\to\infty}\bigg[2^nu\bigg(\frac{t}{2^{n-1}}\bigg) \bigg]=2t\lim_{n\to\infty}\left[\frac{u\bigg(t2^{-(n-1)}\bigg) - u(0)}{t2^{-(n-1)}} \right] = 2tu'(0)
\end{equation}
Thus, we see that
\begin{equation}
    u(2t) = 2tu'(0)
\end{equation}
which means
\begin{equation}
    v(t) = v(0) + tu'(0)
\end{equation}
and hence
\begin{equation}
    w(t) = e^{v(0) + 2tu'(0)} = e^{v(0)}(e^{u'(0)})^{t} = A\rho^t
\end{equation}
as required. \QED
\end{document}